\pdfoutput=1
\documentclass[twocolumn,aps,prb,showpacs,letterpaper]{revtex4-1}

\usepackage{amsmath,amssymb}
\usepackage{bm}
\usepackage{textcomp}
\usepackage{lmodern}
\usepackage[pdftex]{graphicx}

\newcommand{\SRO}{$\text{Sr}_{\text{2}}\text{RuO}_{\text{4}}$}

\begin{document}

\title{Absence of $^{\textrm{17}}$O Knight-Shift Changes across the
First-Order Phase Transition Line in
$\textrm{Sr}_{\textrm{2}}\textrm{RuO}_{\textrm{4}}$}

\author{Masahiro Manago}
  \email{manago@scphys.kyoto-u.ac.jp}
\author{Kenji Ishida}
   \email{kishida@scphys.kyoto-u.ac.jp}
\author{Zhiqiang Mao}
  \altaffiliation{Present address:
  Department of Physics, Tulane University, New Orleans, LA}
\author{Yoshiteru Maeno}
  \affiliation{Department of Physics, Graduate School of Science,
  Kyoto University, Kyoto 606-8502, Japan}

\date{\today}

\begin{abstract}
We performed $^{17}$O nuclear magnetic resonance measurements on
superconducting (SC) $\textrm{Sr}_{\textrm{2}}\textrm{RuO}_{\textrm{4}}$ under
in-plane magnetic fields.
We found that no new signal appears in the SC state and that the $^{17}$O
Knight shifts obtained from the double-site measurements remain constant across
the first-order phase-transition line, as well as across the second-order
phase-transition line as already reported.
The present results indicate that the SC spin susceptibility does not decrease
in the high-field region, although a magnetization jump in the SC state was
reported at low temperatures.
Because the spin susceptibility is unchanged in the SC state in
$\textrm{Sr}_{\textrm{2}}\textrm{RuO}_{\textrm{4}}$, we suggest that the
first-order phase transition across the upper critical field should be
interpreted as a depairing mechanism other than the conventional
Pauli-paramagnetic effect.
\end{abstract}

\maketitle

The layered perovskite \SRO\ has attracted special attention, because it has
been suggested that \SRO\ may be a chiral $p$-wave spin-triplet
superconductor.\cite{JPSJ.81.011009}
The chiral state is shown from the broken time-reversal symmetry probed by
\textmu SR\cite{Nature.394.558} and Kerr-effect\cite{PhysRevLett.97.167002}
measurements.
The existence of spin-triplet equal-spin pairing is based on experimental
results that show the spin susceptibility is unchanged on passing through the
superconducting (SC) transition temperature $T_\text{c}$, as revealed by
nuclear magnetic resonance (NMR) Knight-shift measurements at the Ru and O
sites\cite{Nature.396.658,JLTP117.1587,PhysRevB.63.060507,%
PhysRevLett.93.167004,JPSJ.76.024716} and polarized neutron scattering
measurements.\cite{PhysRevLett.85.5412}
The chiral $p$-wave spin-triplet state would be an SC state
analogous to the superfluid $^3$He A-phase with two dimensionality.

However, several recent experimental results are difficult to interpret
with the above SC state.
The first-order (FO) SC-normal (S-N) transition%
\cite{PhysRevLett.110.077003} accompanied by a clear magnetization jump%
\cite{PhysRevB.90.220502} is observed in a low-temperature region near the
upper critical field $H_\text{c2}$ for fields parallel to the $ab$ plane.
This abrupt S-N transition suggests that \SRO\ is a spin-singlet
superconductor, because this cannot be interpreted by the conventional orbital
depairing effect but seems to be explained consistently by the conventional
Pauli-paramagnetic effect.
Indeed, the experimental $\mu_0 H_\text{c2}$ for $T \rightarrow 0$ nearly
matches the Pauli-limiting field $\mu_0 H_\text{Pauli}$ estimated using the
well-known formula $\mu_0 H_\text{Pauli} = [2\mu_0 E_\text{cond}/
(\chi_\text{n} - \chi_\text{sc})]^{1/2} \sim 1.4$ T with $\chi_\text{sc} = 0$,
where $E_\text{cond}$ is the SC condensation energy and $\chi_\text{n}$ and
$\chi_\text{sc}$ are the spin susceptibilities in the normal and SC states,
respectively.
Here, $\chi_\text{sc}$ = 0 means that the spin susceptibility totally
vanishes in the SC state, which contradicts the above
spin-susceptibility results showing $\chi_\text{n} = \chi_\text{sc}$.

Recently, we performed $^{99}$Ru Knight-shift ($^{99}K$) measurements
again to re-examine the previous results, and found a new phenomenon
that the spin susceptibility slightly \emph{increases} in the
SC state at lower magnetic fields.\cite{PhysRevB.92.100502}
We reported that this experimental result further suggests the
spin-triplet equal-spin pairing state.\cite{JPSJ.83.053701} 
Because the hyperfine coupling constant $A_\text{hf}$ at the $^{99}$Ru site is
largest among the nuclei that are feasible for NMR in \SRO\
($^{99}A_\text{hf} \sim -25$ T$/\mu_\text{B}$),\cite{PhysRevB.63.060507}
the shift of the $^{99}$Ru-NMR spectrum is also largest.
Thus, the $^{99}K$ measurement is suitable for detecting tiny changes of
the spin susceptibility $\Delta \chi_\text{s}$ through the change of the Knight
shift $\Delta^{99}K$ using the relation of
$\Delta^{99}K \propto A_\text{hf} \Delta \chi_\text{s}$.
However, if the $^{99}$Ru NMR signal arising from the SC fraction appears far
from that of the normal state owing to the large $\Delta \chi_\text{s}$ and
is much weaker than the latter signal, it might be possible that we have not
detected the SC signal owing to the poor signal-to-noise ratio.
Indeed, such a separation between the normal- and SC-state signals
was observed in the FO region in CeCoIn$_5$,%
\cite{PhysRevLett.97.227002,JPhysConfSeries.150.052135}
and the SC-state NMR signal in the FO region is weaker than the normal-state
signal depending on the temperatures and fields.

To exclude this possibility, we performed $^{17}$O-NMR measurements to take
advantage of the NMR intensity of $^{17}$O being roughly a hundred-times
larger than that of $^{99}$Ru.
Because $A_\text{hf}$ of the planar O site is one order of magnitude smaller
than that of Ru, as discussed later, the separation between the signals
of the normal and SC states in the case of spin-singlet pairing is not large,
and both the signals can be recorded by a Fourier-transformed spectrum
at one frequency.
In addition, the previous $^{17}$O-NMR Knight-shift measurements%
\cite{Nature.396.658,JLTP117.1587} were mainly performed at lower magnetic
fields below 1.1 T to avoid the suppression of the superconductivity by the
field.
In this study, we focused on the $^{17}$O Knight shift mainly in the field
range of the FO transition, and also measured the Knight shift across the S-N
transition driven by field rotation at low temperatures.
We found that no new signal appears in the SC state, and that the $^{17}$O
Knight shift exhibits no anomaly even across the FO transition line, as well as
in the lower-field region.
These can exclude the possibility that the $^{99}$Ru Knight shift decreases
in the SC state, and suggest that the electrons form triplet-pairing just below
$H_\text{c2}$ even at low temperatures, where the superconductivity is strongly
suppressed by magnetic fields.

\begin{figure*}
	\centering
	\includegraphics{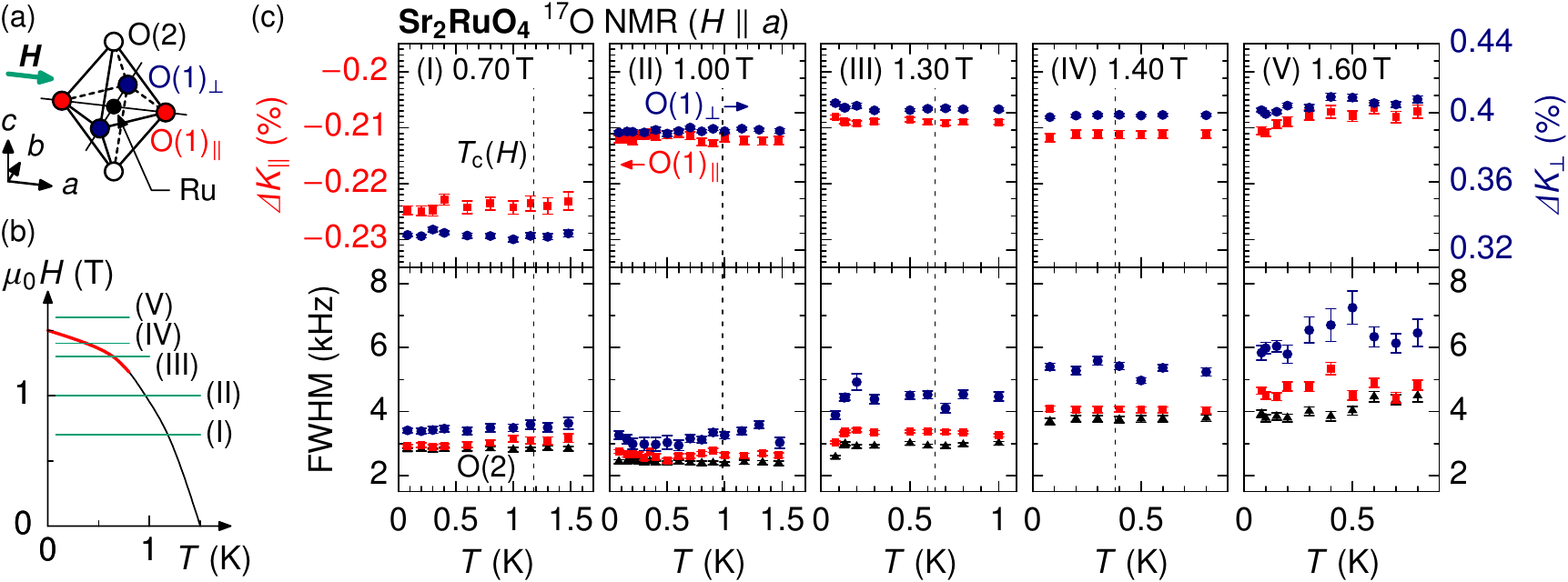}
	\caption{\label{fig:knightshift}(Color online)
	(a) A part of the unit cell of \SRO, indicating the oxygen sites
	and the field direction.
	(b) The field and temperature range of the present Knight-shift
	measurement on the $H$--$T$ phase diagram of \SRO, based on
	Ref.~\onlinecite{PhysRevLett.110.077003}.
	The thick solid boundary represents the first-order transition line.
	(c) (Top) The $^{17}$O Knight-shift differences,
	$\Delta K_{\parallel,\perp} = K_{1\parallel,\perp} - K_2$,
	in \SRO\ under $H \parallel a$ at the central lines
	($1/2 \leftrightarrow -1/2$).
	The square (circle) symbol denotes $\Delta K_{\parallel}$
	($\Delta K_{\perp}$), corresponding to the left (right) axis.
	The vertical dashed lines represent $T_\text{c}(H)$ obtained by
	ac susceptibility measurements using an NMR coil.
	The Knight shift is analyzed within the second-order perturbation
	with respect to the nuclear quadrupole interaction,
	and the apparent field dependence of $\Delta K_{\parallel,\perp}$
	is due to the higher-order terms.
	This does not affect our conclusion.
	(Bottom) The full widths at half maxima (FWHMs) of the $^{17}$O lines.
	The black triangle symbol represents the O(2),
	and the others are the same as in the Knight shifts.}
\end{figure*}

We performed the $^{17}$O NMR measurements on an $^{17}$O-enriched single
crystalline \SRO\ with $T_\text{c} \sim 1.5$ K.
\SRO\ has two inequivalent O sites, O(1) in the RuO$_2$ plane and the
apical O(2) in the SrO plane, as shown in Fig.~\ref{fig:knightshift}(a).
The O(1) signal splits into two lines, O(1)$_{\parallel}$ and
O(1)$_{\perp}$, in the magnetic field along the $a$ axis, where the $\parallel$
($\perp$) symbol denotes the O site with the magnetic field parallel
(perpendicular) to the Ru-O-Ru bonds.
Thus, three distinct NMR central lines are observed as shown later.
The hyperfine coupling constant of the planar O(1) nucleus with the
electronic spins is larger than that of the apical O(2) nucleus
($^{17}A_\text{hf}^{\parallel} \sim -1.9$ T/$\mu_\text{B}$,
$^{17}A_\text{hf}^{\perp} \sim 2.7$ T/$\mu_\text{B}$, and
$^{17}A_\text{hf}^\text{apical} \sim 0.2$ T/$\mu_\text{B}$).\cite{JPSJ.67.3945}
We examined the Knight-shift difference between the O(1) and O(2) sites,
because the Meissner diamagnetization in the SC state
and the small drift of the applied magnetic field, which can be macroscopic
variations, are eliminated by this subtraction, and a precise Knight-shift
measurement can be performed.
The difference of the Knight shift is expressed as
\begin{eqnarray}
	\Delta K_{\parallel,\perp} &\equiv& K_{1\parallel,\perp} - K_2 \notag \\
	&=& (A_{\text{hf}}^{\parallel,\perp} - A_{\text{hf}}^{\text{apical}})
	\chi_\text{s} + \text{const.}
	\label{eq:DeltaK}
\end{eqnarray}
Here, $K_i$ are the Knight shifts at the O($i$) sites ($i=1,2$),
and $\chi_\text{s}$ is the spin susceptibility.
Multiple spin components were introduced for the spin part
of the Knight shift in this system.\cite{PhysRevLett.81.3006}
This will be discussed later in this paper.
The constant term corresponds to the orbital shift, which is usually
temperature-independent and small in the $^{17}$O nucleus.%
\cite{PhysRevLett.81.3006}
The O(2) line can be approximately regarded as a reference signal of the
internal magnetic field owing to the smaller hyperfine coupling constant than
that of the O(1) site.\cite{JPSJ.67.3945}
Such a double-site Knight-shift measurement was also performed in the previous
$^{99}$Ru NMR.\cite{PhysRevB.92.100502}
The temperature-sweep NMR measurements were performed in various magnetic
fields as shown in Fig.~\ref{fig:knightshift}(b).
The Meissner-shielding signal was also measured to confirm that the NMR
measurements were indeed performed in the SC state and to detect some anomalies
related to the FO transition nature.

The temperature dependence of the Knight-shift differences defined by
Eq.~(\ref{eq:DeltaK}) is summarized in Fig.~\ref{fig:knightshift} (c).
The Knight shifts remain constant at both the first- and second-order phase
transition regions.
The lower-field results reproduce the previous results,%
\cite{Nature.396.658,JLTP117.1587} and the detailed higher-field results
of O constitute the new information obtained in this study.

\begin{figure}
	\centering
	\includegraphics{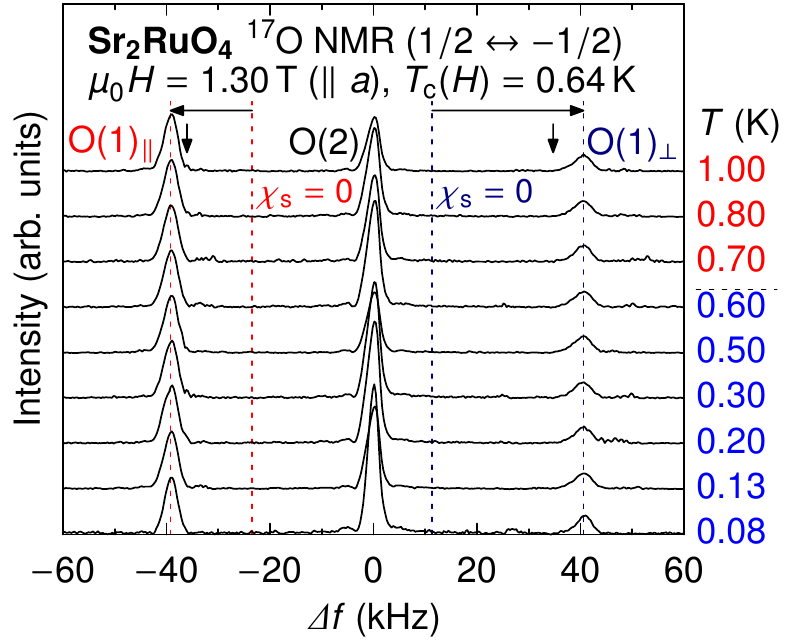}
	\caption{\label{fig:spectra}(Color online)
	The $^{17}$O-NMR spectra of \SRO\ at 1.30 T parallel to the
	$a$ axis at three central lines ($1/2 \leftrightarrow -1/2$)
	observed simultaneously.
	The horizontal axis represents difference of the frequency from the
	apical O(2) line at each temperature.
	The above horizontal arrows represent the total Knight shifts
	of the O(1)$_{\parallel,\perp}$ from the frequencies corresponding
	to $\chi_\text{s} = 0$, which arise from the multiple bands
	of \SRO.\cite{PhysRevLett.81.3006}
	The vertical arrows indicate the expected line positions for
	the spin-singlet pairing (see text for details).
	}
\end{figure}
The spectra at 1.30 T are shown in Fig.~\ref{fig:spectra}.
The spin part of the Knight shifts at the O(1) sites are shown with
the horizontal arrows.\cite{JPSJ.67.3945,PhysRevLett.81.3006}
If a spin-singlet pairing is realized in \SRO, roughly 20\% of $\chi_\text{s}$
decreases at 1.3 T, as estimated from the specific-heat measurement under
the magnetic field.\cite{JPSJ.83.083706}
This can be detected for the present resolution, because the
frequency changes are recognizable quantities as shown by the vertical arrows.
No line shift or new lines were detected in the SC state.
This is the main result of this paper, and indicates that $\chi_\text{s}$
is unchanged even in the field-region of the FO transition.

In the recent $^{99}$Ru-NMR measurement at lower fields, a tiny increase of the
spin susceptibility was reported as mentioned above.\cite{PhysRevB.92.100502}
However, no clear increase of the spin susceptibility was detected in the
present $^{17}$O measurement.
The additional $\sim 2$\% spin polarization,\cite{PhysRevB.92.100502} which
was detected by the $^{99}$Ru-NMR measurement, corresponds to the
$\Delta f \sim 0.2$ kHz line shift in the O(1)$_{\parallel}$ line.
Because the estimated shift is comparable to the present frequency resolution
of $\sim 0.3$ kHz, the absence of a clear increase would be reasonable.

We performed the NMR measurement carefully to avoid any heating of the sample
by the NMR rf pulse fields.
The sample was immersed in $^3$He-$^4$He mixture to avoid any rf heating.
Nevertheless, rf heating becomes recognizable with decreasing temperature,
and thus the rf pulse-power-dependence of the NMR spectra were measured to
inspect the rf heating effect.
No clear power dependence of the NMR spectra was detected at 0.13 K at 1.30 T
in the SC state when the rf pulse power was reduced to $1/8$ of the
ordinary level with a fixed pulse width of 7 \textmu s (not shown here).
Although the pulse power cannot be made arbitrarily small owing to the weak NMR
intensity with weaker rf pulse fields, destruction of the superconductivity
is unlikely to occur in the power range used in this study.
Thus, we conclude that the unchanged Knight shift is not caused by
the rf heating, but an intrinsic property of \SRO.

\begin{figure}
	\centering
	\includegraphics{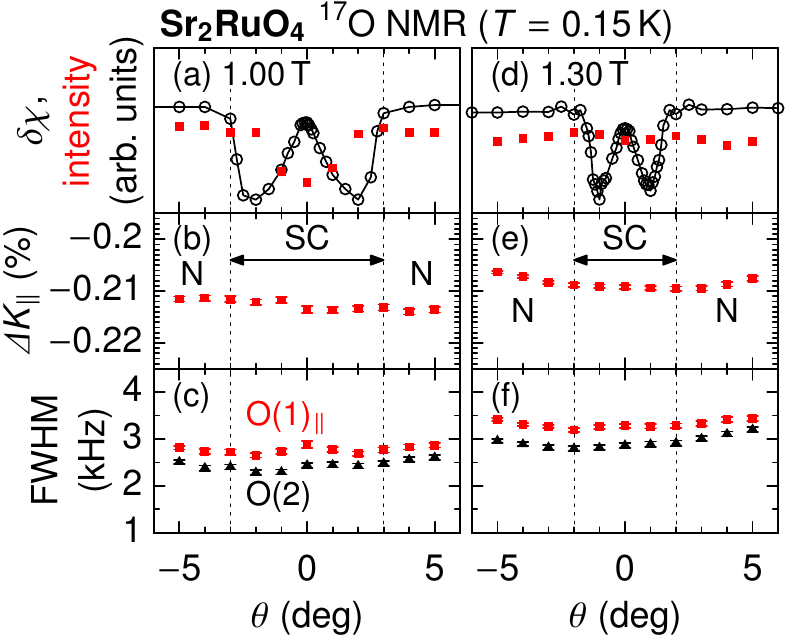}
	\caption{\label{fig:angle_Knightshift}(Color online)
	(a, d) Field-angle dependence of the Meissner shielding
	of \SRO\ at 1.00 and 1.30 T at 0.15 K measured by an NMR coil
	(open circles with line).
	The ac field is parallel to the $\mathit{ab}$ plane and
	perpendicular to the static field, and its frequency is
	$f \sim 7$ MHz.
	$\theta$ is the angle of the static field from the $a$ axis toward the
	$c$ axis.
	The dashed lines represent the SC-normal transition lines
	determined from the Meissner shielding.
	The NMR intensity at the O(1) site, reflecting the shielding of the
	electromagnetic field in the SC state, is also shown (closed squares).
	The lower horizontal axes are the origin for the NMR intensity.
	(b, e) $\theta$ dependence of the Knight-shift differences
	of the O(1)$_{\parallel}$ sites.
	(c, f) The FWHMs of the O(1)$_{\parallel}$ and O(2) lines.
	The black triangle symbol represents the O(2).
	All the NMR spectra were obtained at fixed NMR pulse conditions.
	}
\end{figure}

The field-angle dependence of the Knight shift was measured at 0.15 K.
The field immediately exceeds $H_\text{c2}$ by tilting the applied
field from the $\mathit{ab}$ plane, and thus the Knight shifts in the SC and
normal states can be compared at a fixed temperature with the same NMR pulse
conditions.
There was no clear change of the Knight shift or broadening observed in the
SC state at 1.00 and 1.30 T as shown in Fig.~\ref{fig:angle_Knightshift},
where $\theta$ is the angle between the applied field and the
$\mathit{ab}$ plane.
This result suggests that the spin susceptibility is unchanged even across
the S-N transition induced by the small tilt of the applied field.

The field-angle dependence of the shielding effect shown in
Fig.~\ref{fig:angle_Knightshift}, obtained by an ac field parallel to the
$\mathit{ab}$ plane, has a double-peak structure, and becomes weak
where the field is in the $\mathit{ab}$ plane.
This dependence is understood as the suppression of the Meissner shielding,
which is characteristic of the quasi-two-dimensional supercurrent of the SC
\SRO, as discussed in Refs.~\onlinecite{JPSJ.76.024716} and
\onlinecite{JPSJ.65.2220}.
Because the line width is independent of the strength of the shielding,
the diamagnetic field is considered to be extremely small%
\cite{Nature.396.658,PhysRevLett.111.087003} and cannot be detected by the
nuclear spins with relatively small gyromagnetic ratios.
However, at 1.00 T, the NMR intensity at the O(1) site in the SC state
decreases to about $1/2$ that of the normal state; this is because the
shielding of the rf field is larger in the SC state than in the normal
state.
This indicates that the present NMR spectra are indeed obtained in the SC
state.

Although the signs of the superconductivity were detected in the field-region
of the FO transition, features of the FO transition were not observed in the
present sample with the $^{17}$O NMR and the Meissner-shielding measurements.
Because the present sample has much larger mass ($ \sim 70$ mg) than those used
in the previous study detecting the FO transition,\cite{PhysRevLett.110.077003}
it may be difficult to detect it in the present sample even by another methods
such as the specific-heat or magnetization.\cite{PhysRevB.90.220502}
However, we can safely say that the present sample is a high-quality sample
with $T_\text{c} \sim 1.5$ K and as good as the samples showing the FO
transition.

One may consider that the unchanged Knight shift is a consequence of the
cancellation of the Knight-shift changes in different orbitals because \SRO\ is
a multiband system.
We analyze the Knight shift in the SC state following the discussion by Imai
\textit{et al.}\cite{PhysRevLett.81.3006}
In their model, the $^{17}$O Knight shifts are expressed by the spin parts of
the different Ru $4d$ and O $2s$ electrons, because the spin polarization in
the Ru $4d_{xy}$ and $4d_{zx}$ orbitals is transferred to the O $2p_y$ and
$2p_z$ orbitals, respectively, owing to the covalency of $\pi$ bonds.
The orbital Knight shifts of the O sites are assumed to be negligibly small,
and a nearly temperature-independent isotropic component is ascribed to the
O $2s$ electrons.
The O $2p$ spin parts have anisotropic dipole symmetry:
the dipolar field takes a maximum value along the lobe of the $2p$ orbital and
$-1/2$ of the maximum value along the two orthogonal directions.
The spin part of in-plane components of the O(1) Knight shifts are
then expressed as\cite{PhysRevLett.81.3006}
\begin{eqnarray}
	K_{1, \parallel} &=& \frac{1}{N_\text{A} \mu_\text{B}}
	(-C \chi_{xy} - D \chi_{zx} + \sigma \chi_{2s}),\\
	K_{1, \perp}     &=& \frac{1}{N_\text{A} \mu_\text{B}}
	(2C \chi_{xy} - D \chi_{zx} + \sigma \chi_{2s}),
\end{eqnarray}
where $N_\text{A}$ is Avogadro's number, $\chi_{xy}$, $\chi_{zx}$, and
$\chi_{2s}$ are the partial spin susceptibilities of $4d_{xy}$, $4d_{zx}$, and
$2s$ electrons, respectively, $C$ and $D$ are the dipolar hyperfine coupling
constants of the $2p$ electrons, and $\sigma$ is the isotropic $2s$ hyperfine
coupling constant.
It is possible to set the coupling constants so that the spin part of the
Knight shift is canceled for either the O(1)$_{\parallel}$ or O(1)$_{\perp}$
site, but not for both O(1) sites.
Specifically, the constant $^{17}$O Knight shifts for both O(1)$_{\parallel}$
and O(1)$_{\perp}$ imply that at least $\chi_{xy}$, which arises from the
quasi-two-dimensional (Q2D) $\gamma$ band, does not decrease in the SC state
because $\chi_{xy} \propto K_{1, \perp} - K_{1, \parallel}$.
This multiband treatment makes it clearer that the $^{17}$O Knight shift
provides important information on the band-dependent electronic spin
susceptibility.

Although we cannot rule out the possibility that $\chi_{zx}$ and $\chi_{2s}$
decrease in the SC state while keeping the Knight shifts unchanged,
it is unlikely to occur for all measurement fields.
We note that the Ru Knight-shift value is large and negative,%
\cite{PhysRevB.56.R505,PhysRevLett.81.3006,PhysRevB.63.060507}
because negative isotropic core polarization of the $4d$ orbitals is dominant
and positive $5s$ spin contribution is small.
Thus, in the Ru Knight shift the cancellation does not occur among the
different orbitals.
Therefore, it is also suggested from the constant (or even increasing) Ru
Knight shift that all the components of the $d$-electron spin
susceptibilities does not decrease in the SC state under the magnetic field.

Because the present results indicate that the spin susceptibility is unchanged
in the SC state, it is necessary to consider a depairing mechanism other than
the Pauli-paramagnetic effect in the magnetic field.
In this case, we first need to assume that the spin part does not strongly
contribute to the free energy under the magnetic field.
One possibility is that the Cooper pair is formed between electrons in
different layers by interlayer coupling, and the superconductivity is
destroyed owing to suppression of the interlayer coupling by the external
field parallel to the $\mathit{ab}$ plane, because the coherence length
along the $c$ axis is somewhat longer than the interlayer spacing but of a
similar magnitude.\cite{PhysRevB.91.104514} 
Although it is unlikely that the interlayer interaction is dominant in \SRO\
because the conductivity is two dimensional, the three dimensionality might be
crucially important for understanding the depairing mechanism
on the superconductivity in \SRO.

Another important aspect to consider is the strong spin-orbit coupling in
\SRO, as pointed out by Haverkort \textit{et al.}\cite{PhysRevLett.101.026406}
The $k$-dependent orientation of the expectation value of the spin strongly
mixes the spin-singlet and spin-triplet pairings as seen in the superconductors
with inversion symmetry breaking.\cite{JPSJ.76.051008,JPSJ.83.061014}
However, the Knight shift corresponding to the spin-singlet component should
decrease in the SC state.
Thus, even in this case, the unchanged Knight shift suggests that the
spin-triplet component is dominant in \SRO.

Quite recently, Ramires and Sigrist have pointed out the importance of the
inter-orbital effect in multi-orbital superconductors under magnetic
fields.\cite{2016arXiv160503827}
When the magnetic field is applied along the $\mathit{ab}$ plane, the energy
gain arising from the orbital polarization in the normal state could overcome
the SC condensation energy.
This mechanism could lead to the suppression of the SC phase, and would be able
to explain why the $H$--$T$ phase diagram of \SRO\ is similar to the one
where the Pauli-paramagnetic effect is present.
We also speculate that the presence of multiple bands with different magnetic
properties might be crucial: it is well known that incommensurate
antiferromagnetic (AFM) fluctuations are present,\cite{PhysRevLett.83.3320}
which arise from the Fermi-surface nesting between quasi-one-dimensional (Q1D)
$\alpha$ and $\beta$ bands, and the fluctuations are close to magnetic
instability.
In contrast, strong AFM fluctuations do not exist in the Q2D $\gamma$ band.
If the triplet superconductivity arises from the $\gamma$ band, the
superconductivity of \SRO\ would be immediately destroyed by the AFM
fluctuations induced in the $\gamma$ band when the coupling between
Q1D and Q2D bands becomes stronger under the in-plane magnetic field.
Further studies are required to clarify this possibility.

In summary, we found that no new signal appears in the SC state by
precise $^{17}$O-NMR Knight-shift measurements in the SC \SRO,
and that the spin susceptibility is unchanged across the FO transition line
as well as across second-order one.
Because the present and previous studies suggest that the Pauli-paramagnetic
effect is absent in this system, an alternative depairing mechanism in the
magnetic field is necessary.

The authors thank S. Yonezawa, Y. Yanase, and A. Ramires for valuable
discussions.
This work was supported by Kyoto University LTM Center, and by Grant-in-Aids
for Scientific Research (Grants No.~JP15H05745 and No.~JP26287078),
Grant-in-Aids for Scientific Research on Innovative Areas ``J-Physics''
(Grants No.~JP15H05882, No.~JP15H05884, and No.~JP15K21732), and
``Topological Materials Science'' (Grant No.~JP15H05852) from
the Japan Society for the Promotion of Science.

\end{document}